\documentclass[preprint2]{aastex}

\slugcomment{To appear in AJ}

\shorttitle{Adaptive optics observations of quasar hosts}
\shortauthors{Lacy et al.}

\begin{document}


\title{Observations of quasar hosts with adaptive optics 
at Lick Observatory}


\author{Mark Lacy \altaffilmark{1,2,3}, Elinor L.\ Gates \altaffilmark{4}, 
Susan E.\ Ridgway \altaffilmark{5}, Wim de Vries \altaffilmark{2}, 
Gabriela Canalizo \altaffilmark{2}, James P.\ Lloyd \altaffilmark{6}, 
James R.\ Graham \altaffilmark{6}} 

\altaffiltext{1}{SIRTF Science Center, MS 220-6, California Institute of 
Technology, 1200 E.\ California Boulevard, Pasadena, CA 91125; 
mlacy@ipac.caltech.edu}

\altaffiltext{2}{IGPP, L-413, Lawrence Livermore National Laboratory, 
Livermore, CA 94550 and Department of Physics, University 
of California, 1 Shields Avenue, Davis, CA 95616; wdevries, 
canalizo@igpp.ucllnl.org}

\altaffiltext{3}{Department of Physics, University of California, 
1 Shields Avenue, Davis, CA 95616}

\altaffiltext{4}{Lick Observatory, P.O. Box 85, Mount Hamilton, CA 95140; 
egates@ucolick.org}

\altaffiltext{5}{Department of Physics and Astronomy, 
Johns Hopkins University, 3400 N.\ Charles Street, Baltimore, MD 21218;
ridgway@pha.jhu.edu}

\altaffiltext{6}{Department of Astronomy, University of California, 
601 Campbell Hall, Berkeley, CA 94720; jpl,jrg@astron.berkeley.edu}

\begin{abstract}
We present near-infrared $H$-band
observations of the hosts of three $z\sim 1$ quasars from the
Sloan Digital Sky Survey made
with the  adaptive optics system at Lick Observatory. 
We derive a PSF for each quasar 
and model the host plus quasar nucleus to obtain magnitudes and approximate
scale sizes for the host galaxies. We find our recovered host galaxies
are similar to those found for $z\sim 1$ quasars 
observed with the {\em Hubble Space Telescope}. They also have, 
with one interesting exception, black hole mass estimates from their bulge 
luminosities which are consistent with those from emission-line widths. We 
thus demonstrate that adaptive optics can be successfully used for the 
quantitative study of quasar host galaxies, with the caveat that better
PSF calibration will be needed for studies of the hosts of significantly 
brighter or higher redshift quasars with the Lick system. 
\end{abstract}



\section{Introduction}

All massive elliptical galaxies, and all the bulge components of spiral 
galaxies, 
seem to contain black holes whose masses correlate well with the depth of 
the potential wells of the stellar systems containing them (Ferrarese \& 
Merritt 
2000; Gebhardt et al.\ 2000). This implies that all such systems were probably
quasars at some stage in their lifetimes, and indeed the luminosity density 
produced by quasars is probably consistent with this (Yu \& Tremaine 2002).
The importance of the evolution of quasars as diagnostics of galaxy formation 
models was recognized by Kauffmann \& Haehnelt (2000), who produced 
predictions for the nature of quasar host galaxies as a function of quasar 
luminosity and redshift, based around semi-analytic models of galaxy 
formation.

Two main problems exist with current studies of quasar hosts. First, the 
samples are small. The hosts of only $\sim 30$ 
quasars with $z\sim 1-3$ have been observed with 
NICMOS on the {\em Hubble Space Telescope} (HST) 
(Ridgway et al.\ 2001; Kukula et al.\ 2001; Rix et al.\ 2001). Comparison
with, e.g., the Kauffmann \& Haehnelt models using samples of this size
is difficult, 
as the models predict a wide range in host galaxy magnitudes for a given
quasar luminosity and significant redshift evolution in the host 
magnitudes. 
Second, quasar samples are selected using very different 
techniques, with different selection biases. For example, UV excess selection 
may select against spiral hosts, which are dustier than ellipticals, and radio
selection probably selects the biggest black holes, in giant elliptical 
hosts (e.g.\ Lacy et al.\ 2001). Studying large samples of high 
redshift quasar hosts with HST is impractical
due to the large amount of time required. Therefore our best hope of 
improving our knowledge of the nature of quasar hosts at 
$z\stackrel{>}{_{\sim}} 1$ is to use adaptive optics (AO) from the ground. 
So far, only small samples (such as the one in this paper) have been studied 
with AO. 
The quasar survey component of the Sloan Digital Sky Survey (SDSS), however,
contains
a large enough number of quasars that a significant 
number of quasars have good AO guide stars nearby. Furthermore, because this
survey selects quasars on the basis of their being unresolved optically and  
and having colors lying off the stellar locus (Fan 1999; Richards et al.\ 
2001), it is much more sensitive than most previous optical surveys to 
quasars with a modest amount of dust reddening. We thus expect that host
galaxies of quasars from this survey should give a more
complete picture of the quasar host galaxy population.

AO is, however, is a new technique 
with problems of its own. Previous attempts at studying host galaxies 
with AO,  although largely successful in detecting both close companions and
diffuse emission around the quasars, have
encountered problems with PSF characterization and stability, 
which has limited the amount of quantitative information 
obtainable from the images 
(Stockton, Canalizo \& Close 1998; Hutchings et al.\ 1999; M\'{a}rquez 
et al.\ 2001). 
In this paper we present a pilot study of a small 
sample of $z\sim 1$ quasars from the SDSS Early Data Release (Stoughton 
et al.\ 2002; hereafter EDR) 
observed with the Lick Adaptive Optics system in Natural Guide Star mode.
We discuss issues related to PSF stability and subtraction, and present the 
results and a brief discussion of our observations and prospects for future 
quantitative studies of larger samples of quasar hosts with AO. We assume
a cosmology with $H_0=65 {\rm kms^{-1}Mpc^{-1}}$, $\Omega_{\rm M} = 0.3$ and
$\Omega_{\Lambda} = 0.7$

\section{Sample selection}

The targets were selected from $0.7<z<1.5$ quasars in EDR list by 
matching them to the  
HST Guide Star Catalog, then selecting objects with guide stars
bright enough to act as natural guide stars for the Lick Adaptive
Optics System ($R \stackrel{<}{_{\sim}}12$) within 45-arcsec. 
The three quasars in this paper were then selected on the
basis of RA and guide star suitability. They were
SDSS~J005008.48+011330.4 ($z=1.14$; hereafter SDSS~0050+0113), 
SDSS~J172732.39+584634.4 ($z=0.84$; hereafter SDSS~1727+5846) and
SDSS~J232351.60+004034.4 ($z=0.76$; hereafter SDSS~2323+0040). Only 
SDSS~1727+5846 is radio-loud. Table 1 gives details of the quasars,
and Table 2 the details of the quasar guide star
(hereafter QGS) chosen for each of them.

\section{Observations}

\subsection{The Lick AO system}

Data were acquired using the Natural Guide Star AO 
system on the 3-m Shane Telescope at Lick Observatory. 
The AO system has 40 subapertures and 
uses a Shack-Hartmann wavefront sensor and a 
61-actuator deformable mirror. The Lick AO system is further
described in Bauman et al.\ (1999) and Gavel et al.\ (2000).  
The IRCAL near-infrared
camera was used for the observations. This camera contains a PICNIC
HgCdTe 256x256 array (Lloyd et al.\ 2000), which sits behind the AO system
reimaging optics.  The f/28 output of the AO system gives a plate scale of
0.076\arcsec /pixel and a field of view of 19.4\arcsec $\times$ 19.4 \arcsec.

\subsection{Observational strategy}

In addition to a small PSF, surface brightness sensitivity is essential for 
successful observations of quasar hosts. The best AO performance of the Lick
system is obtained in $K$-band. However, the thermal background in $K$ is 
high due to the warm optical elements in the AO system. 
We therefore chose to make our observations in $H$-band.

We tried to pick an observational strategy that was a good compromise between
low overheads and reliable PSF monitoring. Observations of the quasars were
therefore interleaved with on-axis observations of their guide stars
at intervals of $\approx 45-75$ min. We also made a single 
observation of a PSF star - PSF guide star 
pair for each quasar, selected such that the PSF star was
at the same distance and position angle from its guide star as the quasar was 
from its guide star. 
We used the PSF star observation to 
calibrate the effect of going off-axis (i.e.\ the anisoplanatism). The 
PSF star -- PSF guide star (hereafter PGS) pairs are listed in Table 3.

The quasar observations were made on the night of 2001 August 14th (UT). The
data were taken as several five-point mosaics, with each pointing 
lasting 3-5 min, depending on sky stability. Small offsets of 1-2 arcsec were
made between each mosaic. Total integration times were 75min for 
SDSS~0050+0113, 135 min for
SDSS~1727+5846 and 120 min for SDSS~2323+0040.
The interleaved observations of 
the QGS were made on the same five-point grid with exposure times of 
10s per frame. The PSF stars were also observed on a five-point grid, with
exposure times of 20-60 s per point. 
Flux calibration was achieved using the standard
star HD162208 (Elias et al.\ 1982). The natural seeing was $\approx 0.75$ 
arcsec in $H$-band. AO corrections were made at 100 Hz for all the targets. 
This maintained a level of $\sim 200$ counts on each subaperture
of the wavefront sensor for our guide stars. Images of the
fields are shown in Figure 1. Typically we were able to achieve FWHM of 
about 0.2 arcsec on axis (compared to the diffraction limit of 
0.14 arcsec), however this degraded to about 0.4 arcsec off axis.

\section{Analysis}

Following flat-fielding, the quasar data were analysed with the {\sc dimsum} 
package within {\sc iraf}. Just before the final combination step, a second
order polynomial was subtracted from each column to improve the 
the images, removing a ``step'' between two halves of the array. The 
most likely cause of this seems to be different, time-variable bias 
offsets which were not well-removed by the dark subtraction.

\section{PSF synthesis}

The AO PSF is both variable in time and dependent on many of 
the observing parameters. The brightness and color 
of the guide star (which determines the accuracy of the AO 
correction), the anisoplanatism (which may vary both with guide-star -- object
distance and position angle between the objects and the guide star), and the 
color of the object compared to that of the PSF star
all need to be considered. 
Ideally, one would interleave quasar observations with frequent
sampling of the off-axis PSF, using observations of a nearby PSF star
-- PGS pair well matched in guide star brightness, color, separation 
and position angle to the quasar -- QGS pair. To produce a more efficient 
procedure, however, we observed as described in section 3.2, and 
subsequently attempted to 
reconstruct the PSF from frequent observations of the on-axis guide star, and 
a single observation of a PSF star -- PGS pair chosen for proximity to the 
quasar rather than for exact matches in color and brightness to the 
quasar -- QGS pair. (We have not attempted to
correct for color differences between the PSF and quasar.)
Our procedure, as described below,
enables us to 
synthesize a PSF with a FWHM close to that of the quasar 
observations, even if the QGS and PGS have 
different brightnesses and therefore AO corrections which differ in quality.
It also allows us to remove any component of the aberration due to 
anisoplanatism effects which is constant in time. Steinbring et al.\
(2002) show that a similar strategy of determining a kernel map 
for the off-axis variation of the PSF by observing a crowded stellar 
field can be effective, and remove the bulk of the anisoplanatism effects
on the PSF.

The first step involves deconvolving the PSF star observation by the 
PGS. The Lucy deconvolution algorithm was used. Sufficient iterations were
made to reduce the FWHM of the PSF star to significantly less than that of 
its guide star, with care being taken to stop the deconvolution process
before obvious artifacts (e.g.\ ringing) appeared in the 
deconvolved image. This produces an ``off-axis kernel'' (Figure 2).
In the second step, the off-axis kernel is then
convolved with the average of the on-axis guide star observations to
produce a synthesized PSF. This technique minimizes the problems of a 
mis-match in brightness and color of the PGS and QGS.  
\onecolumn

\begin{table}
\caption{The quasar sample}
{\footnotesize
\begin{tabular}{lccccc}
\tableline
\tableline
Quasar &RA & Dec &$z$ & $r_{\rm AB}$ &$S_{1.4{\rm GHz}}$\\ 
       & (J2000) & (J2000)  &   & & (mJy)   \\
\tableline
SDSS 0050+0113& 00 50 08.48&  +01 13 30.40 &1.14 &18.8 &$<1$\\ 
SDSS 1727+5864& 17 27 32.41&  +58 46 34.44 &0.84 &18.4& 216 \\ 
SDSS 2323+0040& 23 23 51.60&  +00 40 34.36 &0.76 &20.0 &$<1$\\ 
\tableline
\end{tabular}
}
\end{table}

\begin{table}
\caption{Guide stars}
{\footnotesize
\begin{tabular}{lccccc}
\tableline
\tableline
Quasar &QGS RA& QGS Dec&QGS $R$-mag& PA& QGS Distance\\ 
       & (J2000)   & (J2000) &      &    &(arcsec)\\
\tableline
SDSS 0050+0113&00 50 05.75&  +01 13 27.4&10.5&266&41\\ 
SDSS 1727+5864&17 27 34.45& +58 47 14.7 &9.6 &22&43\\ 
SDSS 2323+0040&23 23 54.12&  +00 40 14.6&10.3&118&42\\ 
\tableline
\end{tabular}
}
\end{table}

\begin{table}
\caption{PSF stars and PSF Guide Stars}
\footnotesize{
\begin{tabular}{lccrcrcccr}
\tableline
\tableline
Quasar  & 
\multicolumn{2}{c}{PSF Star} & \multicolumn{2}{c}{PSF Star Guide Star}& PA & 
Separation\\ 
       & RA(2000) Dec(2000)& $R$-mag & RA(2000) Dec(2000)& $R$-mag && (arcsec) \\ 
\tableline
SDSS 0050+0113&00 56 54.74 +00 44  2.3 & 15.7 &00 56 52.19 +00 43 58.5 &11.3&264&42\\
SDSS 1727+5864&17 18 33.16 +59 30 12.3&14.0& 17 18 35.01 +59 30 55.3 &  9.9&18 &45\\ 
SDSS 2323+0040&23 24 17.60 +01 41 40.2&12.1&23 24 20.21 +01 41 29.5 & 10.6 &105&41\\
\tableline
\end{tabular}
}
\end{table}

\begin{table}
\caption{Image quality}
\footnotesize{
\begin{tabular}{lccccc}
\tableline
\tableline
Quasar & Raw QSO FWHM &Synthesized PSF FWHM & QGS FWHM & PSF Star FWHM & PGS FWHM\\
       &(arcsec)      &(arcsec)             &(arcsec)  &(arcsec)       &(arcsec)\\
\tableline
SDSS 0050+0113&0.36& 0.38 & 0.20    & 0.39 & 0.23 \\
SDSS 1727+5846&0.44& 0.44 & 0.18    &  0.49  & 0.18\\
SDSS 2323+0040&0.45& 0.41$^{*}$& 0.20 & 0.29 & 0.14\\
\tableline
\end{tabular}

\noindent
$^{*}$ convolved to account for registration errors, 0.37$^{''}$ prior
to convolution. All FWHM are from Gaussian fits. }
\end{table}

\begin{figure}
\centering
\includegraphics[scale=0.9]{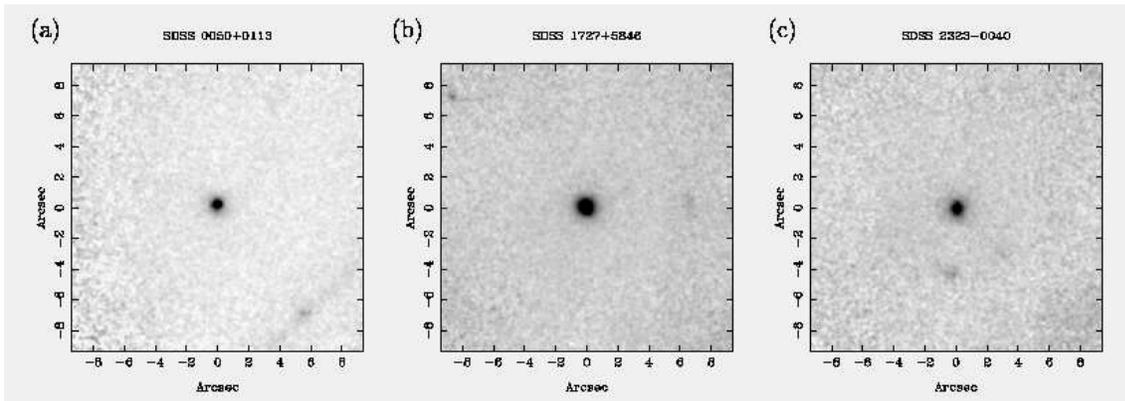}
\caption{(a) The field of SDSS~0050+0113, smoothed with a Gaussian of 
FWHM 0.18$^{''}$. The diagonal stripe to the lower-right is an 
artifact, caused by scattered light. (b) The field of SDSS~1727+5846, 
smoothed with a Gaussian of FWHM 0.18$^{''}$. (c) The field of 
SDSS~2323-0040, smoothed with a Gaussian of FWHM 0.18$^{''}$. }
\end{figure}

\begin{figure}
\centering
\includegraphics[scale=0.9]{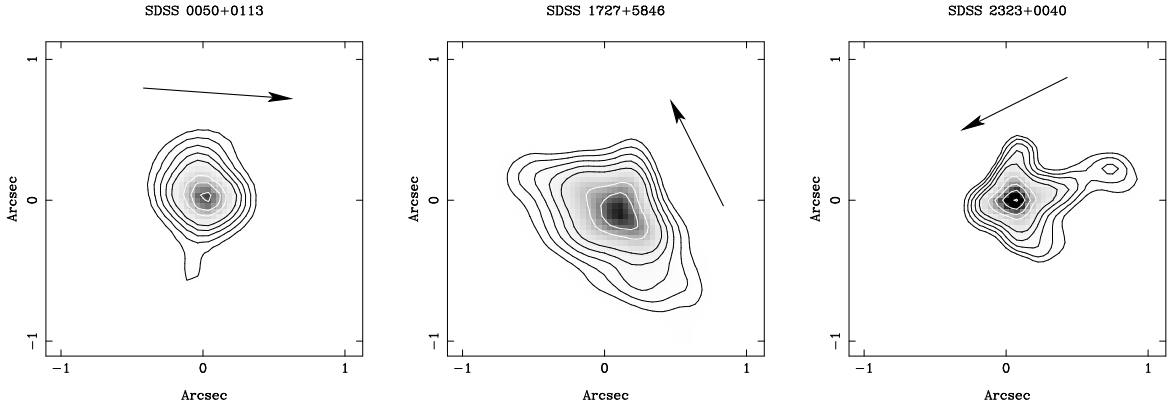}
\caption{The off-axis kernels for each of our observations, obtained 
by deconvolving the PSF star by the PGS observations. This kernel
is convolved with the QGS observations to synthesize the 
observed PSF. The arrows indicate the approximate direction of the guide 
stars. Note that the kernel tends to be elongated in this direction (see, 
e.g.\ McClure et al.\ (1991)). 
The contours are 
logarithmic, spaced by factors of two, overplotted on a linear 
greyscale stretch.}
\end{figure}

\begin{figure}
\centering
\includegraphics[scale=0.9]{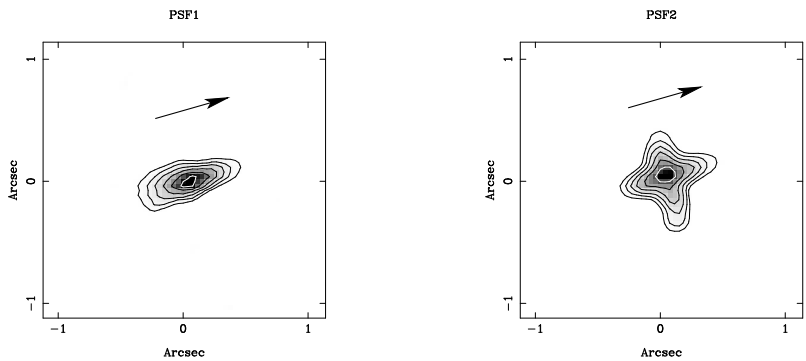}
\caption{The off-axis kernel for two different PSF stars with similar 
separations and position angles to their guide stars, observed 
3hrs apart. The arrows indicate the approximate direction of the guide 
stars. The contours are 
logarithmic, spaced by factors of two, overplotted on a linear 
greyscale stretch.}
\end{figure}

\begin{table}
\caption{Observed host galaxy properties}
{\footnotesize
\begin{tabular}{lccccccccccccc}
\tableline
\tableline
Quasar & $m_{f}$ &$r_{\rm in}$& $m_{e}$ & $m_{d}$  &$\Delta m$&$\theta_{f}$ & $\theta_{e}$ & $\theta_{d}$&$\Delta \theta$& $\chi^2_{e}/dof$& $\chi^2_{d}/dof$
& Axial ratio & Host\\
       &         & &         &          &          &    &       
&             &               &                  & && PA\\
\tableline
SDSS0050+0113&18.8&0.38& 18.2   &  18.5 & 0.3 & 0.75 & 0.7    & 0.4 &0.3  & 1.13 &1.12&1.5&57\\
SDSS1727+5846&19.0&0.38& 18.6   &  18.8 &0.25  & 1.2 & 1.2    & 1.1&0.3  & 0.89 &0.89&1.0&...\\
SDSS2323+0040&18.6&0.15& 18.6   &  18.9 &0.15  & 0.6 & 0.4    & 0.5&0.1  & 0.93 &0.94&1.3&0\\
\tableline
\end{tabular}

\noindent
Notes: $m_f$ is the 
magnitude measured by subtracting the inner aperture to flatness, and
$r_{\rm in}$ the radius of the inner aperture in arcseconds. 
$m_e$ is the magnitude of the best-fitting elliptical galaxy model and 
$m_d$ that of the best-fitting disk galaxy model. 
All magnitudes are $H$-band, Vega magnitudes, and are measured in square
apertures 7.5-arcsec on a side centered on the quasar.
$\Delta m$ is the estimated error in the magnitudes (both systematic and 
random). $\theta_f$, $\theta_e$ and $\theta_d$ are the angular half-light
radii for the subtracted-to-flatness, elliptical model and disk model 
hosts respectively, all in arcseconds.
$\Delta \theta$ is the estimated error in the angular half-light radius
in arcseconds.
$\chi^2_e/dof$ and $\chi^2_d/dof$ are the reduced $\chi^2$ values for the 
elliptical and disk model fits, respectively. The axial ratio and host 
PA are measured from the outer isophotes of the subtracted-to-flatness
hosts.}
\end{table}

\begin{table}
\caption{Derived quasar host properties using the elliptical host 
model.}
{\footnotesize
\begin{tabular}{lcccccccc}
\tableline
\tableline
Quasar & $M_{V}$ & $r_{1/2}$ & $M_{\rm BH}$ (Host) &$M_{\rm BH}$ (BLR)& $L/L_{Edd}$ \\
       &         & (kpc)     & ($M_{\odot}$)   &($M_{\odot}$)&\\
\tableline
SDSS0050+0113&-23.4&6$\pm 3$&$8.3\times 10^8$&$7.0\times 10^8$&0.3-0.4 \\
SDSS1727+5846&-22.2&10$\pm 2$&$3.5\times 10^8$&$2.0\times 10^8$&0.5-1.0 \\
SDSS2323+0040&-21.8&3.1$\pm 0.5$&$2.9\times 10^8$&$4.0\times 10^7$&0.14-1.0 \\
\tableline
\end{tabular}

\noindent
Notes: $r_{1/2}$ is the half-light radius, $M_{\rm BH}$ (Host) is the 
black hole mass estimated from the host galaxy luminosity,
$M_{\rm BH}$ (BLR) is the same quantity estimated using the width of the
Mg{\sc ii} emission line. $L/L_{\rm Edd}$ is the luminosity of the quasar
in Eddington units, calculated assuming a bolometric correction of 12
to rest-frame $B$-band (Elvis et al.\ 1994).}
\end{table}

\begin{figure}
\includegraphics[scale=1.0]{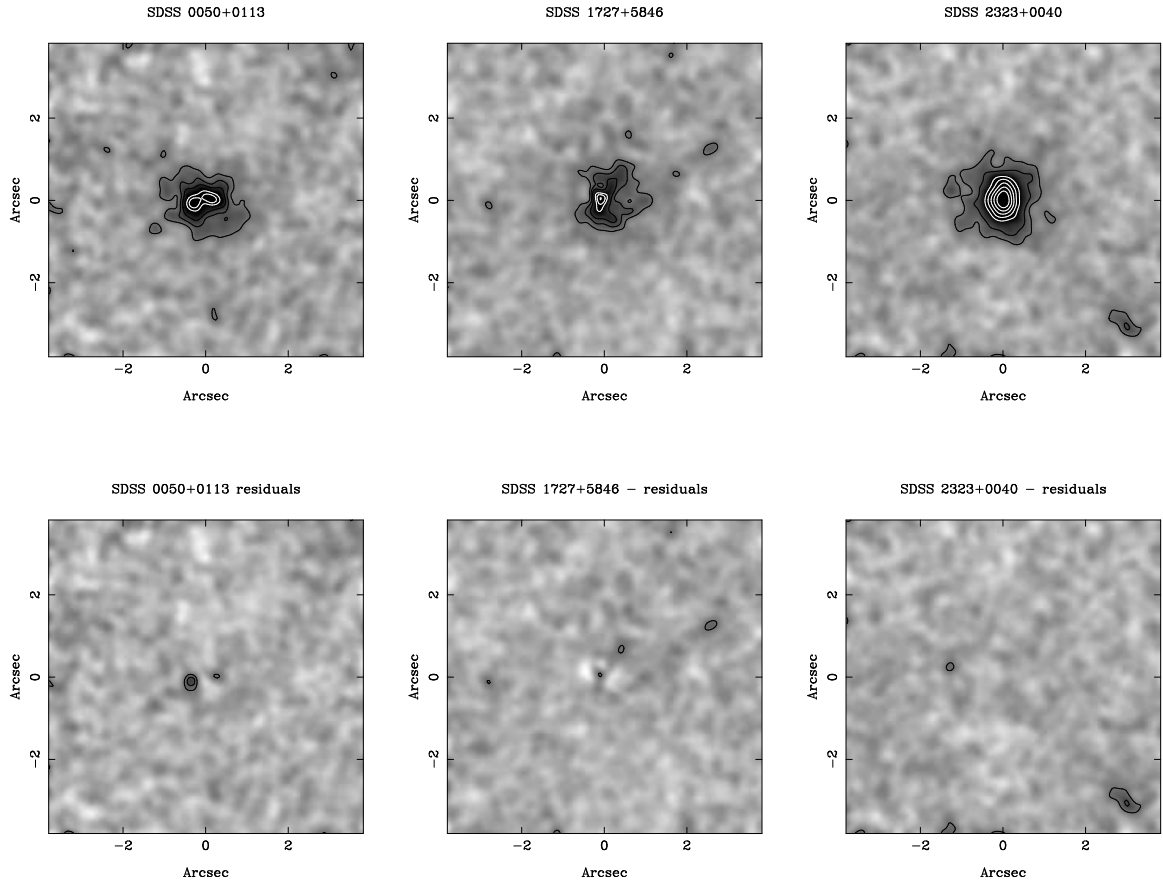}
\caption{The top row shows the three host galaxies, as revealed after 
subtraction of 
the best-fitting point source model of the quasar nucleus. The bottom 
row shows the residuals after subtraction of the elliptical 
host galaxy model. 
All the images have
been smoothed by a 0.26\arcsec$\;$FWHM Gaussian, and the greyscale and 
bottom contour levels are scaled to the image noise in the same way in 
each plot. North is up and east to the
left.}
\end{figure}

\twocolumn

To investigate the variation of the off-axis kernel with time and the 
particular choice of PSF star/PGS pair, 
we performed a test using data obtained during non-photometric
conditions on 2002 May 26 (UT). We used the same observational setup and 
observed in similar seeing conditions as our 2001 August observations. This  
test involved observing two different PSF stars close to 1709+6225 (J2000).
The two PSF star/PGS pairs had similar separations (about 40\arcsec) 
and position angles, and the observations were separated in time by 3~hrs.
The results of this test showed that the off-axis kernel was
significantly different, although there was a common axis of symmetry aligned
with the vector joining the PSF to its guide star (Figure 3). The guide star
for the second PSF was about two magnitudes brighter than that for the 
first PSF, so the AO corrections may have differed in quality. 
More tests need to be made to establish which of 
time variability of the off-axis kernel or AO correction quality
are more important, but this test at least allowed us to obtain some estimate 
of the uncertainty in the reconstructed PSFs.

In one case (SDSS~2323+0040) the quasar was sufficiently 
faint that inaccuracies in individual image registrations when constructing 
the final mosaic add significantly to the width of the effective PSF. 
Fortunately, the quasar is relatively faint compared to the host galaxy
in this case, so the uncertainty in the exact amount by which the PSF was 
broadened translates to only a small error in the host galaxy magnitude.  
We estimated the uncertainty in the centroiding by comparing the centroid 
estimates
used to combine the data with independent estimates using the {\sc iraf}
task {\sc radprof} and estimates based on the peak pixel position. 
Both techniques suggested
that the rms error in the centroiding was 0.05-arcsec in $x$ and $y$, so we
convolved synthesized 
PSF by a Gaussian with $\sigma=0.05$ arcsec. 
Table 4 summarizes 
the QGS, PGS and PSF star FWHM. 

\section{Determination of host galaxy properties}

\subsection{Model-independent host properties}

Before making a full model of the host plus quasar system, we
obtained approximate constraints on the host 
magnitudes and scale-sizes by subtracting the PSF, scaled such that the 
residual was approximately flat in the center of the quasar host within a 
radius $r_{\rm in}$, and declined monotonically outside $r_{\rm in}$. 
The radius $r_{\rm in}$ was chosen such that outside this 
radius the  
PSF residuals were small compared to the noise in the images. 
This technique, although somewhat subjective and 
resulting in a small oversubtraction of the nuclei, has the advantage that 
galaxy parameters are model independent. It also provides a good starting 
point for the full modeling, described below. 

\subsection{Modeling the host galaxies}

The host galaxies were modelled following the procedure in 
McLure, Dunlop \& Kukula (2000) by fitting 
PSF plus galaxy model profiles (convolved by 
the PSF) by minimizing $\chi^2$. Systematic errors from the 
PSF subtraction dominate the error budget close to the center; to prevent
these regions dominating the fit, the inner 0.1-0.4 arcsec region was
downweighted by factors of 0.5-0.7 to ensure that the $\chi^2$ surface 
was fairly uniform across the fitting aperture. We fitted the
scale size and flux of the host galaxy model, and the flux and 
position of the nucleus. To obtain stable convergence, we fixed the 
host galaxy position angle and axial ratio using measurements of the outer
isophotes of the ``subtracted to flatness'' hosts discussed above.
Small adjustments in the background level were also found to be necessary at 
this stage to set the residual to zero at large radius.  
The host galaxies, after subtraction of the best-fitting PSF corresponding
to the quasar nucleus, are shown in the top row of Figure 4, with 
the residuals after subtraction of the galaxy models shown below. 
Both elliptical and disk models
were tried, but neither was a significantly better fit to the data. The 
results of the modeling are summarized in Table 5.

Errors in the magnitudes of the hosts were estimated by combining estimated 
errors from noise, systematics from PSF mismatch, and photometric errors.
Errors in the magnitudes and scale sizes due to noise in the $\chi^2$ 
fitting were estimated by a bootstrap 
technique which involved randomly shuffling the pixel values in the residual
image, adding back the model and refitting with a starting vector randomly 
changed by $\sim 10$\% in each fitted parameter (PSF scale, galaxy flux and 
half-light radius) from the best-fitting parameter set. This exercise was 
repeated 30 times for each quasar to estimate the error in the 
host galaxy magnitudes and scale sizes.

To estimate the size of the likely systematic error due to PSF mismatch,
we tested the modeling code with a PSF rather than 
quasar data. 
We took the PSF for SDSS~2323+0040 and convolved it to the 
width of the PSF for SDSS~1727+5847.
We then used the same modeling code as for the real quasars to fit to the 
PSF for SDSS~1727+5847 as if it were a quasar plus host galaxy system, 
using the convolved PSF of SDSS~2323+0040 as the PSF in the model. 
Constraining the half-light radius of the residual ``host galaxy'' 
to be between 0.1 and 1.2 arcsec, we estimated that
errors due to PSF mis-match in this case were $\sim 10-20$\%. 
We also performed a similar test using our two PSFs from 
2002 May 26 (UT). Again we scaled the 
results to SDSS~1727+5847, finding a residual with $H=20.5$, i.e.\ a 17\%
flux error.  
The errors for the monotonic subtraction technique were smaller ($\sim 10$\%),
as expected, as this estimate is less dependent on the nature of the PSF 
residuals close to the center of the host.

The error estimates for the quasar magnitudes were then constructed by 
adding in 
quadrature the estimated photometric error of 0.1 mag., the errors from the
bootstrap modeling and an estimated error from PSF mis-match (0.2~mag.\ for
SDSS~0050+0113 and SDSS~1727+5846, where the quasar flux dominates, 
and 0.1~mag.\ for SDSS~2323+0040 where the 
PSF and host fluxes are comparable).
The errors for the scale sizes are those from the bootstrap procedure; as 
these scale sizes are most sensitive to regions of the fit away from the 
center, they are probably fair estimates.

\subsection{Galaxy magnitudes and black hole masses}

To calculate the absolute magnitudes of the host galaxies, and thereby 
obtain an estimate of their black hole masses by using the black hole 
mass -- bulge luminosity relation, we first
needed to estimate the $K$-corrections from observed-frame $H$-band
to rest-frame $V$-band. In order to do this, we assumed the 
host galaxies were ellipticals which formed at high redshift, $z\sim 5$. 
This is consistent with most known quasar hosts at low redshifts, and in any 
case the difference in $K$-correction between a spiral and an elliptical
host is only $\sim 0.2$ magnitudes. We will also assume the host galaxy 
parameters from the elliptical galaxy models (though again these are little 
different from those of the other models).
The assumption that our hosts are ellipticals was more important 
when we tried to estimate the black hole masses from the galaxy luminosities; 
naturally we will have overestimated the 
black hole masses for disk-dominated systems if we assume our hosts 
are ellipticals. The example of
low redshift AGN hosts suggests, however, 
that even those in spiral galaxies are mostly in 
early-type spirals, with high bulge fractions (Dunlop et al.\ 2001). 

We used the elliptical galaxy model of Fioc \& Rocca-Volmerange (1997)
to make the $K$-corrections. To calculate the 
black hole masses, we also needed to estimate the passive evolution of the
stellar luminosity of the host galaxy in order to relate the host
galaxies at $z=1$ to those in the local Universe, on which the black 
hole mass -- bulge luminosity correlation is calibrated. 
Observations suggest elliptical 
galaxies fade by about one magnitude between $z\approx 1$ and the present
(e.g.\ Lubin \& Sandage 2001). We again used the Fioc \& Rocca-Volmerange 
predictions for their elliptical galaxy model, which results in a 0.9~mag.\ 
correction in $V$-band at $z=1$.

Black hole masses were estimated from the galaxy luminosities using the 
empirical relationship of van der Marel (1999), adapted to 
$H_0=65 \;{\rm kms^{-1}Mpc^{-1}}$. Masses were also 
estimated using the FWHM of the Mg{\sc ii} emission line as described 
by McLure \& Jarvis (2002). Absolute magnitudes and black hole mass
estimates are given in Table 6.



\section{Comments on individual objects}

\subsection{SDSS~0050+0113}

This host is the most luminous galaxy in the sample, so was fairly 
easily detected despite its high redshift and 
the relatively short total integration time we obtained. Some PSF residual 
is visible in Figure 4, but extended host emission is clearly detected.

\subsection{SDSS~1727+5846}

This quasar is the only radio-loud quasar in our sample. It is also the 
least well-detected host, with clear PSF residual being visible in Figure 4,
due to a combination of a relatively faint host and a bright quasar nucleus.
The radio structure is slightly resolved in the Faint Images of the 
Radio Sky at Twenty-cm (FIRST) survey, with 
a size of 3\arcsec, a PA of 55 deg and a total flux density of 180 mJy; 
30\arcsec$\;$away at PA 43 deg there is an unresolved 22mJy source, which 
may be associated. The NRAO VLA Sky Survey (NVSS)
flux density is 216 mJy, greater than the sum of the two candidate components
in FIRST, suggesting that the quasar has 
additional diffuse emission which is resolved out in FIRST. The radio 
spectral index is $\alpha \approx -0.4$ from 151~MHz to 1.4~GHz, 
steepening to 
$\alpha \approx -0.7$ between 1.4~GHz and 5~GHz (where $\alpha$ is defined
in the sense that flux density, $S_{\nu} \propto \nu^{-\alpha}$). The radio 
morphology and low-frequency spectral index both suggest that the radio flux 
density is enhanced by Doppler boosting, though probably not by 
a factor of more than a few. Nevertheless, this may help to explain why this 
quasar has a relatively low-luminosity host and small black hole mass for a 
radio-loud quasar. The black hole mass from the emission line width 
may also be underestimated somewhat due to orientation effects if the 
broad-line region (BLR) is in the form of a disk perpendicular to the 
radio axis (Brotherton 1996).  

\subsection{SDSS~2323+0040}

The host galaxy of this quasar was comfortably detected, largely due to the 
weakness of the quasar nucleus, and consequently has the best-determined
magnitude and scale size. It is the only one of our quasar hosts to show
evidence of interaction. There is a faint companion galaxy 1.4\arcsec$\;$east
of the quasar with faint emission between it and the quasar (Figure 4). In 
addition, there are other galaxies nearby, one 4.3\arcsec$\;$to the 
southwest, and one 4.5\arcsec$\;$to the south, 
though there is  no clear evidence of 
interaction of either of these galaxies with the host.

\begin{figure}
\includegraphics[scale=0.4]{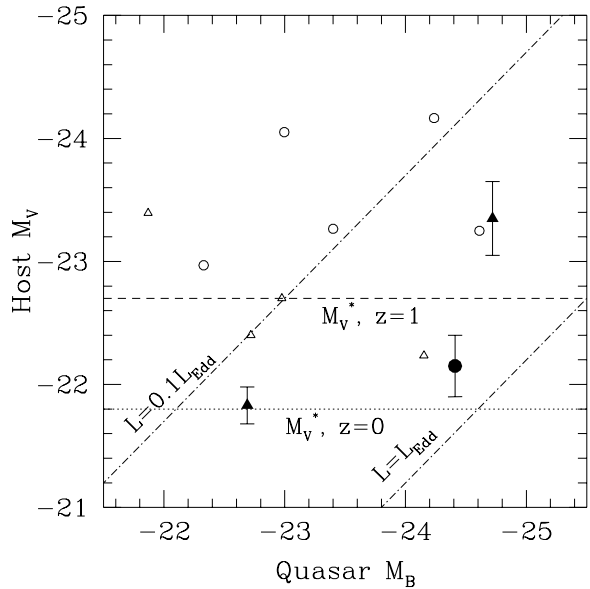}
\caption{Host $M_V$ (elliptical galaxy model) 
plotted against quasar $M_B$. Our points are
shown as filled symbols, the points of Kukula et al.\ (2002) are 
indicated by open symbols. Triangles denote radio-loud quasars, 
circles radio-quiets. The dotted line indicates the 
magnitude of an $L^*$ galaxy at $z=0$, the dashed line the same galaxy
at $z=1$, assuming the passive evolution of a stellar population which 
formed at $z\sim 5$. The two dot-dashed lines show the positions of quasars
accreting at the Eddington rate and one-tenth of the Eddington rate.}
\vspace*{1.0in}
\end{figure}



\section{Discussion}

Our AO observations have proved successful at detecting hosts around three
$z\sim 1$ quasars. The accuracy of our host galaxy flux density 
measurements are, as expected, limited most by the accuracy of our PSFs. 
For our $z\sim 1$ quasars, errors in the PSF have
not been large enough to prevent us obtaining scientifically-useful results. 
It will, however, be necessary to make a better PSF calibration 
if we wish to study systems with significantly brighter quasars relative 
to their hosts, or higher redshift quasars where surface brightness dimming 
makes extended emission from the host harder to detect. This may simply 
involve taking more frequent PSF observations, or better matching the 
PSF star/PGS pair to the quasar/QGS pair, but
it may turn out that better PSF calibration will require a different
approach than the one described here. 

Although a substantial improvement over the natural seeing at Lick was 
obtained through the use of adaptive optics, our resultant image quality is 
only comparable to the natural seeing at a first-class astronomy site such as 
Mauna Kea. Similar techniques could, however, be used at such sites to 
deliver even better quality images. One complication not addressed in this 
paper is that of altitude-azimuth mounted telescopes. In these, the component
of the PSF due to the telescope optics rotates 
with the field, and thus will vary rapidly with time, though the off-axis 
component, determined by the atmosphere, should not be affected. How useful 
our technique would be for
such telescopes will depend on the details of the optical system, although
given that a large contribution to the PSF width comes from the off-axis 
component our technique may still be useful.

The magnitudes and scale sizes of the quasar hosts are 
comparable to the 
nine $z\sim 1$ quasar hosts from the HST/NICMOS study of Kukula et al.\ (2001)
(Figure 5), although our hosts are on average a little fainter. They are
therefore  
closer to the predictions of Kauffmann \& Haehnelt (2000). Though our sample 
is too small to make a definitive statement, if this trend continues to be 
seen in a larger sample, it may be a result of the SDSS quasar survey having 
fewer selection biases than conventional optical surveys. Previous optical 
surveys tended to select very blue quasars and are thus sensitive to small  
amounts of reddening in the host. One might therefore expect them to 
favour quasars in less dusty hosts, such as giant ellipticals.

A comparison between the black hole mass estimates from the galaxy 
luminosities and the Mg{\sc ii} linewidths shows that 
for SDSS~1727+5946 and SDSS~0050+0113, the agreement is very good, 
within a factor of two, but for SDSS~2323+0040 
the black hole mass estimated from
the linewidth is an order of magnitude less than that derived from the host.
Dunlop et al.\ (2002) also find that their black hole mass estimates for 
$z\sim 0.2$ quasars imaged with HST usually 
agree well, but a few objects also have order-of-magnitude
discrepancies between black hole
mass estimates, in the same sense, that the emission line estimates are 
too low. They ascribe these to disk-like broad-line regions being seen
face-on. However, it is also true that the three most discrepant black hole
mass estimates in the Dunlop et al.\ sample all show evidence for 
interactions, and include the two most spectacular mergers in their sample. 
SDSS~2323+0040 is also the only one of our hosts to show signs of 
interactions. Perhaps the bulge mass luminosity estimates in merger systems
are too high. Possible reasons for this include a starburst in the merger 
system lowering the mass-to-light ratio of the stellar population, or a delay
between the galaxy merger and the merger of the black 
holes of the two galaxies and the subsequent
accretion of significant amounts of 
matter onto the merged black hole. 
Large differences between black hole mass estimates derived
from emission-line widths and host bulge luminosities may thus 
be indicators of a quasar formed from a recent merger event. 
All our quasars are accreting at or 
below the Eddington rate using either black hole mass estimate.

The advent of large samples
of quasars from the SDSS and the Anglo-Australian 2dF surveys
means that significant
numbers of quasars near bright AO guide stars have already become available. 
Even larger samples will be possible using laser guide stars. The image
quality with laser guide stars will also be better as the AO corrections will 
be made on-axis. Thus we expect to be 
able to form statistically-useful samples of high quality quasar host images
in the near future.

\acknowledgments

We thank Don Gavel for valuable support, and the AO group at Lawrence 
Livermore National Laboratory (LLNL) as a whole for their hard work
on the AO system over the years.
We also thank Michael Gregg for help with the host galaxy fitting software,
and the referee, Alan Stockton, for a helpful report.
This work was partly carried out at the Jet Propulsion Laboratory, 
California Institute of Technology, under contract with NASA, and 
partly under the auspices of the U.S.\ Department of Energy, National
Nuclear Security Administration by the University of California, 
LLNL, under contract No.\ W-7405-Eng-48, 
with additional support from NSF grant AST-98-02791 (University of California,
Davis).
The Guide Star Catalog was produced at the Space Telescope Science Institute 
under U.S. Government grant.


\clearpage








\end{document}